\documentclass[fleqn,10pt]{wlscirep}
\usepackage[utf8]{inputenc}
\usepackage[T1]{fontenc}
\usepackage{lscape}
\usepackage{booktabs}
\usepackage{fnpct}
\title{Atomist or Holist? \\ A Diagnosis and Vision for More Productive Interdisciplinary AI Ethics Dialogue}

\author[1,*]{Travis Greene}
\author[2]{Amit Dhurandhar}
\author[1]{Galit Shmueli}
\affil[1]{National Tsing Hua University, Institute of Service Science, Hsinchu, 30013, Taiwan}
\affil[2]{IBM Research, Yorktown Heights, New York, USA}

\affil[*]{corresponding author(s): Travis Greene (travis.greene@iss.nthu.edu.tw)}


\newcommand{\tg}[1]{\textcolor{black}{#1}}

\begin{abstract}
In response to growing recognition of the social impacts of new AI-based technologies, major AI and ML conferences and journals now encourage or require papers to include ethics impact statements and undergo ethics reviews. This move has sparked heated debate concerning the role of ethics in AI research, at times devolving into name-calling and threats of ``cancellation.” We diagnose this conflict as one between atomist and holist ideologies. Among other things, atomists believe facts are and should be kept separate from values, while holists believe facts and values are and should be inextricable from one another. With the goal of reducing disciplinary polarization, we draw on numerous philosophical and historical sources to describe each ideology’s core beliefs and assumptions. Finally, we call on atomists and holists within the ever-expanding data science community to exhibit greater empathy during ethical disagreements and propose four targeted strategies to ensure AI research benefits society.
\end{abstract}
\begin{document}

\flushbottom
\maketitle

\thispagestyle{empty}

\section*{Introduction}

In his 1917 lecture \textit{Science as a Vocation}, Max Weber argues that questions of fact are separate from questions of value \cite{weber1958science}. Given pre-specified ends, the task of science is to find the best means of achieving them. But asking which ends we ought to achieve is a question of values, answerable only by philosophy or religion. Weber was not alone on this point. Eminent mathematicians, physicists, and economists espouse similar views on the separation of science and values. But a growing number of data science researchers see things differently. They argue that values, particularly those related to the desirability of social and political goals, implicitly influence the foundations of data science practices \cite{friedler2021possibility, green2021data, crawford2021atlas}.

As public concern mounts over the use of AI-driven research to enable surveillance technologies, deep fakes, biased language models \cite{weidinger2022taxonomy}, misinformation, addictive behaviors, and the discriminatory use of facial recognition and emotion detection algorithms \cite{Kate2022ProtocolEthics}, data scientists appear divided along ideological lines about what to do. While some may view the inclusion of ethics impact statements in several AI conferences and journals as a sign that the value-neutral ideal of science is no longer tenable, a vocal cadre of data scientists has pushed back against this conclusion, asserting the importance of academic freedom and value-neutrality\cite{DomingosACMOpenLetter}.  Yet the conflict between these two ideological camps threatens to polarize the data science community, lowering the prospects that new AI-based technologies will contribute to our collective well-being.

 This largely historical and conceptual paper is organized as follows. Section \ref{overview} lays out the conflicting ideological views---\emph{atomism} and \emph{holism}---underlying AI ethics debates. Section \ref{rolescientist} contrasts the ``two cultures'' of atomism and holism by presenting a guiding metaphor for each. Section \ref{factval} relates philosophical debate on facts and values to community concerns about ethically evaluating AI-based technologies. Section \ref{statementsocialbenefits} considers how ethics impact statements embody a shift towards a more holist data science culture and discusses the social benefits of value-based discussions. Section \ref{ethicsdemands}, in response, identifies key sticking points of ethics reviews from the perspective of an atomist data scientist. Finally, Section \ref{AccEmp} presents a broad vision for more empathetic discussion of value-related issues and proposes four targeted strategies for more civil discourse.

\section{Atomism vs. holism: An overview of conflicting ideologies}
\label{overview}
 An \emph{ideology} is an all-encompassing world-view advancing a political and ethical vision of the good society  \cite{heywood2021political}. To better understand the nature of ethical disagreements in data science, we introduce a simple ideological taxonomy we call \emph{atomism} and \emph{holism}, drawing on the terminology and work of philosophers\cite{taylor1979atomism, popper2012open, bunge2000systemism, berlin2002liberty} and intellectual historians\cite{skinner2002visions, sowell2002conflict}. At risk of oversimplification, the purpose of our taxonomy is to make each ideology's implicit beliefs, assumptions, and historical foundations more explicit so they can be reflected on, refined, and more openly discussed within the data science community should ethical conflict arise \cite{domingosCancelBlog} (see Table \ref{tab:taxonomy}). In full disclosure, we ourselves have mixed academic backgrounds and self-identify at different points along the ideological spectrum: one is mostly holist, the other mostly atomist, while another is a ``moderate,'' finding aspects of both sides agreeable. We encourage readers to also reflect on where they might fit within this taxonomy.

 To delineate the scope of the arguments and examples, we use the label \emph{data scientist} to cover both industry data scientists who design and deploy socially-impactful AI/ML-based systems, and data scientists---often in academia, but also in industry---whose job descriptions includes publishing research articles in peer-reviewed conferences and journals. Our broad definition includes \emph{AI/ML researchers} and \emph{AI/ML engineers}. We use the terms \emph{data science} and \emph{AI/ML} interchangeably.

\begin{table*}[h]
\caption{Summary of two basic but often conflicting ideologies in the data science community along several core dimensions.}
\label{tab:taxonomy}
\begin{tabular}{|l|l|l|}
\hline
                             & \multicolumn{1}{c|}{\textbf{Atomist}} & \multicolumn{1}{c|}{\textbf{Holist}}      \\ \hline
\textbf{Guiding metaphor}             & Puzzle solver         & Social steward          \\ \hline
\textbf{Facts and values} &
Separate &
Inseparable
\\ \hline
\textbf{Associated ``Isms''} &
  \begin{tabular}[c]{@{}l@{}}(Neo)liberalism, libertarianism, \\ logical positivism, modernism\end{tabular} &
  \begin{tabular}[c]{@{}l@{}}Communitarianism, feminism, \\ post-positivism, post-modernism\end{tabular} \\ \hline
\textbf{Social orientation}           & Individualist             & Collectivist                  \\ \hline
\textbf{Self-concept}                 & Autonomous                            & Relational                                \\ \hline
\textbf{Means of social coordination} & Incentives and markets                & Shared moral values and dialogic exchange \\ \hline
\textbf{Key moral concepts} &
  \begin{tabular}[c]{@{}l@{}}Rights, duties, contracts, \\ impartial justice\end{tabular} &
  \begin{tabular}[c]{@{}l@{}}Empathy, caring, connection, \\ responsivity to vulnerable others\end{tabular} \\ \hline

\textbf{Vision of the good life}      & Neutral and constrained                 & Substantive and unconstrained                  \\ \hline
\textbf{Scientific methodology}       & Data-driven, empiricist, neutral                  & Theory-laden, rationalist, perspectival                \\ \hline
\textbf{Extreme form leads to}        & Technocracy/nihilism/alienation       & Totalitarianism/dogmatism/tribalism       \\ \hline
\end{tabular}
\end{table*}

\subsection{Conflicting visions of self, society, and politics}

 Our diagnosis begins by observing that the conflict between atomist and holist data scientists mirrors philosophical debates between traditional liberals\footnote{We use the term \emph{liberal} in Isaiah Berlin's sense of ``neutral with respect to conceptions of the good,'' \cite{berlin2002liberty} although we are aware it has taken on different connotations in modern-day political discourse.} and communitarians and care ethicists  \cite{slote2007ethics, held2006ethics, noddings2013caring}. The conflict centers on the nature of and relationship between the self and society, the proper balance between social order and personal autonomy \cite{etzioni1996responsive}, and whether a just society should advance any particular view of the good life \cite{sandel1984procedural}. Atomists embody more of an individualist culture---inwardly-focused on his or her individual needs and interests---while holists embody more of a collectivist culture---outwardly-focused on the needs and interests of the social collective \cite{triandis2018individualism}. Atomists and holists thus offer opposing solutions to the basic dilemma facing every society: that of protecting its members from one another without oppressing them\cite{etzioni1996responsive}.  Both atomists and holists claim the opposing ideology---when taken to the extreme---leads to various social and individual pathologies.

Holist data scientists view their identities as relationally embedded in a larger society with a particular and unique history and culture. The relational identity of holists entails various relationships of responsibility and caring for others \cite{noddings2013caring, held2006ethics}. Holists stress the self as concretely situated and constituted through caring relations to particular others. As such, holists believe that a good life and society should foster attentive, caring relations and cultivating empathy for the experiences and suffering of vulnerable others \cite{slote2007ethics, held2006ethics}. Holist data scientists recognize that their technical expertise allows them to understand AI technologies in a way that politicians and ordinary citizens might not. Their privileged knowledge creates an obligation to communicate these concerns publicly so that citizens and policy-makers can act appropriately \cite{anderson2011democracy}. Holists thus see themselves as social stewards or fiduciaries acting on behalf of society, holding an ``unconstrained'' view of the perfectability of human nature and society \cite{sowell2002conflict}. Limiting the freedom of other data scientists in order to protect and prevent harm to vulnerable third parties, marginalized groups, and the environment thus may be a necessary evil in the utopian quest towards the ideal society.

Atomists, as the name implies, tend to see the world reductively \cite{bunge2000systemism}, holding that what separates us is prior to what connects us \cite{held2006ethics}. In political matters, atomists emphasize  separateness and independence from others while focusing on abstract issues of impartial justice, duties and rights. They stress the importance of formal equality under the law and the freedom to enter into voluntary contracts of exchange in markets; they prefer economic systems in which social coordination is achieved through the pursuit of enlightened self-interest, rather than imposed by an external authority \cite{hayek1980individualism}. Atomists---wary of utopia---see the achievement of the good life as tragically constrained by unavoidable trade-offs, limited resources, and human imperfections \cite{sowell2002conflict, crowder2006value}. Finally, atomists value personal autonomy and take violations of their personal integrity seriously \cite{smiley2009moral}. Personal integrity refers to the life projects to which a person is committed \cite{smart1973utilitarianism}. Atomist data scientists feel the imposition of moral responsibility to others unfairly infringes on their freedom to pursue their life projects as they see fit, a view which reflects their individualist view of self and society.

\subsection{Conflicting visions of the role of science in society}

Mirroring economic arguments for specialization and trade, atomists support a gap between fact-based inquiry and value-based inquiry. The job of the technically-trained scientist is to conduct fact-based inquiry and report experimental results. In contrast, the job of policy-makers is to advance the state of society by acting on these results according to  political and ethical values \cite{rudner1953scientist}. How others interpret the facts, and what they do on the basis of these interpretations, is out of the control of the scientist. The atomist thus defends the empiricist position that a judgment of fact entails nothing about value, and vice versa \cite{elgin2013fact}. Factual statements, strictly speaking, do not motivate action or decision \cite{hume2003treatise} (i.e., \emph{ought} cannot be derived from \emph{is}). Atomists argue that intellectual specialization and division of labor---i.e., a logical gap between value-neutral fact finding and value-based decision-making \cite{popper2012open}---is a feature, not a bug, of the arrangement. Hence, atomists see the imposition of ethics impact statements as hampering scientific fact-finding, which is essential to innovation and economic progress.

Atomists are also concerned about how the injection of values can lead to---in the best case---wishful thinking and---in the worst case---to dogmatism and totalitarianism  \cite{popper2012open}.  Atomists cite how Soviet science was notoriously influenced by its interpretation of Marxist philosophy, or how Galileo was forced to recant his support of Copernicanism under the inquisition of the Catholic church.  The dogmatic and totalitarian tendencies of holists were recently summarized by one data science academic who condemned ``the increasing use of repressive actions aimed at limiting the free and unfettered conduct of scientific research and debate" \cite{DomingosACMOpenLetter}. By claiming that science is and ought to be value-free, or at least value-neutral \cite{betz2013defence}, atomists advocate a form of inductive empiricism, tracing back to Francis Bacon and Galileo \cite{lacey2005science}.

Meanwhile, holists advocate a post-positivist attitude to science, where facts and values mutually inform one another \cite{anderson1995value}. They doubt there is a uniquely correct or absolute view of reality \cite{rorty2009philosophy} and emphasize the role of interpretation and perspective \cite{denzin2011sage}. Holists thus reject the value-free ideal of science and the atomist ban on deriving an \emph{ought} from an \emph{is}. Instead, holists believe facts discovered by social/data science ought to be put towards realizing a substantive, utopian vision of the good life and society  \cite{gorski2013beyond, dewey1998essential, sayer2009s}. Free and open discussion of values allows for dialogue aimed at finding overlapping consensus about which values are and should be embodied in data science research and applications. Without a clear articulation of and commitment to shared moral values, new AI technologies may impede the achievement of the good life \cite{floridi2018ai4people}. In particular, holists fear the creation of a faceless and unaccountable \emph{technocracy} narrowly aimed at prediction and control of human behavior resulting from the ``value-neutral,'' instrumental application of science \cite{habermas1985theory}. In short, holists view ethics impact statements as tools for clarifying shared action-guiding values, developing moral character, and inducing democratic deliberation on the responsibility of data scientists to society.

\section{The ``two cultures'' within the data science community}
\label{rolescientist}
In 1959, at the height of Cold War, and as the US military-industrial complex established itself, scientist and writer C.P. Snow worried about a growing divide between two academic cultures---those from the ``hard sciences'' and the ''humanities``---whose specialization rendered them increasingly hostile and unmotivated to communicate with one another\cite{snow2012two}. We suggest a similar dynamic between may be stoking division within the larger data science community. Inspired by philosopher and historian of science Thomas Kuhn, we sketch two guiding metaphors capturing core differences between rival atomist and holist research communities.  

\subsection{Atomists: Data scientists as puzzle solvers}

Through a process of disciplinary socialization and training, atomists see themselves as acquiring a constellation of shared beliefs, values, and techniques---in short, a paradigm---that permits progress on open-problems, or ``puzzles,'' the paradigm identifies as solvable \cite{kuhn1970structure}. \tg{Commitment to the paradigm identifies a researcher as a member of a distinct scientific community.} During periods of ``normal'' science (as opposed to revolutionary science), the legitimacy of the paradigm's values and traditions is presumed, narrowing researchers' focus on the task of more reliably and efficiently gathering relevant facts\cite{kuhn1970structure}. In AI and data science research communities, for instance, shared standards and traditions might take the form of the 0.05 threshold for p-values in statistical hypothesis testing \cite{rudner1953scientist}, or the use of several standard benchmark datasets (e.g., ImageNet, CIFAR-10, Fashion MNIST, etc.) and performance metrics (e.g., precision, recall, or F1 score).  

The chief task of the atomist data scientist is, through sheer creativity and imagination, to make progress in solving the puzzles posed by the paradigm, not to question its legitimacy. Beyond the familiar boundaries of the paradigm there is neither consensus on what a right answer looks like, nor a guarantee that a solution even exists. Atomists concede that while the results of AI research may later be used for what outsiders consider morally good or bad ends, these external judgments ultimately rely on criteria and values foreign to the paradigm. \tg{Because the atomist's scientific identity stems from loyalty to the paradigm, atomists worry that undue focus on external social issues not only slows down puzzle-solving, but threatens both the integrity of the paradigm and their personal autonomy.}

\subsection{Holists: Data scientists as social stewards}
Holist data scientists see the growing public concern over the social impact of AI as anomalous facts signifying a crisis and  ultimately demanding a paradigm shift\cite{kuhn1970structure}.  Holists propose revolutionary changes in perspective and new disciplinary procedures, open-problems, and traditions. In the emergent holist paradigm, data scientists are social stewards\cite{bak2021stewardship} or fiduciaries working on behalf of society and advancing substantive social values and human interests\cite{habermas1971knowledge} through data science \tg{research and applications.} 

The holist conviction that data scientists ought to act as social stewards or fiduciaries is advocated by a growing number of data scientists, particularly in high-stakes healthcare domains \cite{eaneff2020case, stoger2021medical}. The ethical justification of data scientists' fiduciary duty to society derives from their capacity to ``produce consequences that matter to others'' \cite[pg. 110]{goodin1986protecting}. In particular, holists are concerned about unjust power differentials and coercive dependency relationships \cite{held2006ethics} that may arise due to applications of AI-based technologies. The role of social steward or fiduciary aligns with holist beliefs that the self is constituted through social relations and that social responsibilities and caring relations are important for our psychological well-being \cite{sayer2009s}. 

  A fiduciary relation is a special kind of social relation involving  discretion, power, inequality, dependence, vulnerability, trust, and confidence \cite{miller2013justifying}. The word fiduciary itself stems from \emph{fide}, meaning ``trustworthy.''  Fiduciaries have the power and authority to make decisions on behalf of the beneficiary. Common fiduciary relations include trustee-beneficiary, principal-agent, manager-shareholder, lawyer-client, doctor-patient, and parent-child relations. The presumed fiduciary power of holists may be justified on the basis of their unequal access to knowledge of technical details and social applications of emerging AI-based technologies. Given this knowledge asymmetry, fiduciaries abide by a duty of loyalty to act in the best interests of the beneficiary and refrain from opportunism and conflicts of interest \cite{miller2013justifying}. Oriented by a substantive view of the human good, and motivated by a belief in unlimited social progress,
holists feel obligated to ensure AI-based technologies promote human flourishing  \cite{taddeo2018ai}.

\section{The separation of facts and values}
\label{factval}

Atomists and holists disagree about how facts and values relate and their role in data science. To better clarify the fact-value debate, this section examines the issues in more historical detail. At stake is whether values can be subjected to rational scrutiny and thereby provide objective grounds for ethical criticism of AI technologies.

\subsection{The fact-value controversy: A brief historical interlude}
The crux of the fact-value debate centers on whether it is possible to rationally criticize (i.e., by appeal to logic or empirical evidence) both the means used to achieve an end, and the end itself.  Empiricists, following David Hume, tend to argue it is not possible to rationally critique ends; rationalists, following Immanuel Kant, disagree.

 Hume's empiricism was revived in the 20th century in the philosophy of science known as \emph{logical positivism} or \emph{logical empiricism}, which asserted that philosophical problems ultimately stemmed from ambiguities in ordinary language \cite{ayer1952language}. This highly ambitious project aimed to unify all sciences using formal logic and derive experimentally-verifiable hypotheses from formalized versions of scientific theories \cite{rorty1992linguistic}.  Asserting empirical verification as the criterion for meaning was an important methodological move that paved the way for \emph{behaviorism} in psychology, \emph{revealed preference theory} in economics \cite{lewin1996economics}, and \emph{operationalism} in psychometrics \cite{borsboom2005measuring}.

 The focus on empirical verification of scientific statements challenged the idea of objectivity in ethics. What methods of proof or evidence could support ethical judgments? To overcome this problem, some philosophers suggested that ethical judgments were actually composed of two separate components: a descriptive or factual part, and a command-like prescriptive component \cite{hare1991language}. Others concluded that values were emotive or attitudinal in nature and lacking in descriptive, factual content, analogous to cries of pain, groans, or shrieks \cite{ayer1952language}. Thus to say ``X is good'' is nothing more than saying ``hurrah for X!'' or ``I like X; you should too" \cite{stevenson1944ethics}. 
 Logical positivists, in short, claimed moral language is simply not ``about'' anything at all, casting serious doubt on efforts to resolve moral disagreements.

  Despite its onetime popularity, philosophers of science abandoned the logical positivist project as early as the 1960s \cite{suppe2000understanding}. Most philosophers today accept that facts and values are not so cleanly separable \cite{williams2006ethics}. Still, vestiges of the logical positivist project remain influential in other disciplines \cite{putnam2004collapse}, particularly those fond of axiomatic formalization such as economics and computer science.

\subsection{Implications for moral expertise and objectivity }

Evaluating AI-based technologies presumably requires moral expertise. But if the notion of a moral fact is incoherent, how can we make sense of the idea of moral expertise? Does moral expertise imply the existence of objective moral facts, such as "killing is wrong" \cite{elgin2013fact}? Even if we are willing to grant the existence of such facts,  must moral experts not only possess \emph{theoretical}, but also \emph{practical} moral knowledge---the real-world skills needed to correctly apply moral concepts in the right situations \cite{weinstein1994possibility}? Although developing a procedure for testing the validity of moral judgments appears difficult, if not impossible, moral philosophers nevertheless routinely undertake this task \cite{rawls1971theory, habermas1990moral}.

 Another problem is that in most scientific domains, we expect experts to reach consensus on key facts, as it indicates convergence on the truth \cite{williams2006ethics}. Yet moral disagreement is common, and value statements seemingly do not achieve a high degree of consensus, even among experts \cite{steinkamp2008debating}. Nearly all data scientists will agree that \emph{when training a machine learning model, both bias and variance cannot be minimized simultaneously}, but presumably fewer agree with ``value-laden'' statements such as  \emph{a predictive model's accuracy is more important than its interpretability}. 
The lack of consensus around value statements, especially in an era of globalized data science, could be a sign that objective knowledge in the domain of AI ethics may not be possible. Without such objectivity, the legitimacy of publication decisions that include ethical evaluation may be disputed.

Despite these difficulties and the persistent arguments of moral skeptics, we hope to encourage---not  discourage---ethical reflection, especially by those without formal training in philosophy. As AI-based technologies increasingly impact society, AI ethics is more relevant than ever. But a more central role for ethics in AI research means that members of long-estranged academic disciplines will be forced to engage one another, disrupting the current intellectual division of labor\cite{nowotny2005increase} and posing new barriers to communication. Still, we believe the disruption of disciplinary boundaries can benefit society.

\section{Social Benefits of Ethics Impact Statements}
\label{statementsocialbenefits}
 One major step towards acknowledging the ethical issues raised by AI was the decision by the \emph{Neural Information Processing Systems} (NeurIPS) conference in 2020 to introduce a new ethics review process. The process requires submitting authors to include a ``broader impact'' section while a group of ethics experts review papers flagged as problematic by technical reviewers  \cite{ashurst2022ai}. Following the example of NeurIPS, impact statements and ethics reviews by experts are spreading to other conferences and journals \cite{tmlrEthicsstatements}. The \emph{IEEE/CVF Computer Vision and Pattern Recognition Conference} (CVPR) in 2022 adapted the NeurIPS ethics guidelines and ``strongly encourages'' authors to ``discuss the ethical and societal consequences of their work in their papers in a concrete manner'' \cite{cvprConf2022Ethics}. But creating effective ethics impact statements remains challenging. Impact statements are plagued by a variety issues including their complexity, lack of guidance and best practices, lack of purpose, lack of procedural transparency, high opportunity costs, institutional pressure, and various social and cognitive biases affecting authors and referees \cite{prunkl2021institutionalizing}. 

Nevertheless, data scientists can no longer presume their research will have a net positive impact on the world \cite{hecht2021s}. Besides functioning as a kind of layman's guide to new AI-based technology, the inclusion of impact statements is a positive step towards acknowledging the social responsibility of those who design, research, and deploy AI technologies.  Below we identify several ways in which impact statements may benefit data science research and society in general.

\subsubsection*{Train ``citizen'' data scientists and cultivate moral and intellectual virtues}  Writing and critiquing an ethics statement could become a core part of the moral education and training of future data scientists. Impact statements encourage moral reflection and the cultivation of civic and intellectual virtues, such as attentive observation, open-minded imagination, patient reflection, careful analysis, and fair-minded interpretation and assessment \cite{baehr2011inquiring}. The development of moral and intellectual virtues could break down interdisciplinary academic barriers isolating researchers from public debates on AI-based technologies \cite{greene2022barriers}. Hagendorff \cite{hagendorff2020ai}, for instance, suggests cultivating four basic virtues:  justice, honesty, responsibility, and care, along with secondary virtues of prudence and fortitude (speaking truth to power). Impact statements can promote the virtue of humility, too, by requiring data scientists to better understand and communicate the limitations of new AI tools, reducing opportunities for sensationalism in the popular media \cite{johnson2017ai}. To develop these virtues, data science curricula might offer \emph{Embedded EthiCS}\cite{grosz2019embedded} courses to foster discussion on the ethical implications of AI-based technologies and provide training in writing ethics impact statements.

\subsubsection*{Promote disciplinary cross-pollination} 
The big-picture scope of ethics reviews and the diversity of the review groups can contribute to disciplinary cross-pollination. Novel dilemmas, formalizations, and technologies can be brought back and analyzed in different academic communities. These issues can then become part of the ethical training of data scientists. Data science research also confronts legal scholars and regulators with thorny problems of interpretation and application of existing laws, which could lead to better lawmaking. Lastly, cross-pollination can occur when data science researchers experimentally test the assumptions of traditional ethical theories \cite{wallach2008moral}, leading to theory refinement in ethics and moral psychology. For example, deep neural nets can make accurate predictions about the morality of various novel acts (e.g., ``it’s rude to mow the lawn late at night'') as judged by human evaluators \cite{jiang2021delphi, hendrycks2020aligning}.

\subsubsection*{Spur new debates on the purpose and applications of corporate data science research}
Technologists increasingly voice their concerns when corporate applications of technology conflict with their ethical values \cite{davis2021corporate, mitchellFired}. Value-based discourse allows debate on the social and ethical impact of corporate-funded data science applications. One AI-driven application likely to lead to moral disagreement in the data science community is digital advertising \cite{milano2021epistemic}. Jeff Hammerbacher, one of Facebook’s first data scientists, once said, ``The best minds of my generation are thinking about how to make people click ads...That sucks'' \cite{hammerbacherFBads}. Indeed, Google and Facebook are two of the world's largest technology companies employing data scientists and also two of the biggest players in digital advertising. 
Massive industry investment reflects the allocation of vast amounts of data scientist labor to solving problems related to digital advertising. But beyond the obvious economic benefits, what values justify this enormous outlay of human energy and attention?

\subsubsection*{Identify and inspire new forms of data science for social good projects}
 A more transparent debate about values can expand the scope of data scientists' responsibility to new moral communities and stimulate new ideas for data science for social good. Ethics debates can identify new forms of harm before they become systematically entrenched in business practices and research pipelines. 
  Interesting ethical blindspots of AI concern how computer vision, natural language processing (NLP), and other technologies contribute to environmental degradation \cite{crawford2021atlas}, the exploitation of human data labelers---particularly in low-income countries \cite{kshetri2021data} and in NLP crowdsourcing tasks \cite{boaz2021beyond, santy2021use}, and the mistreatment of animals in factory farms \cite{hagendorff2021blind}. 
 Another angle involves developing algorithms to ``boost'' \cite{hertwig2017nudging, lorenz2020behavioural} critical thinking about the quality and veracity of news stories shared on social media in an effort to foster democratic values.

\subsubsection*{Reveal the sociotechnical, political, and value-laden nature of applied data science}
Although one might reasonably argue that we should distinguish between more abstract AI research (e.g., optimization and machine architecture), and more applied work (e.g., education and criminal justice), ethics impact statements can helpfully illuminate the sociotechnical nature of data science. In real-world institutions such as courts or corrections facilities, legal terms and predictive relationships evolve, actors act adversarially, and numerous non-sampling errors can affect algorithmic performance \cite{greene2022forks}. Not only that, many decisions in machine learning pipelines involve ``essentially contested'' and value-laden constructs such as fairness \cite{angerschmid2022fairness} and justice \cite{mittelstadt2019principles}. Sociotechnically-oriented data science recognizes that fairness is more than just a property of an algorithm \cite{selbst2019fairness}: AI can be used as part of a political agenda \cite{green2021data}. One noteworthy example comes from Barabas et al.\cite{barabas2020studying}, who refused to build an algorithm due to concerns about its role in reinforcing structural injustices in the prison system. 


\section{Atomist Concerns Regarding Ethics Statements}
\label{ethicsdemands}
This section considers possible concerns and arguments arising from atomist data scientists.

\subsubsection*{Lack of technical credibility} Multiple cases, including our own personal experience, exist where research is rejected or red flagged for apparent ethical concerns. Sometimes, however, the rejection is unwarranted as the contribution of the paper may be orthogonal or even a mechanism to better understand AI model risks. For example, an explainable AI paper might be rejected because the model it explains (viz. a large language model) has fairness issues. But the model itself is not the contribution. Undesirable behavior uncovered through explanations is not a limitation, but rather a validation of the explanation method. A rejection decision showcases ethics reviewers' inability to understand the work's technical contribution. In larger contexts, AI tools and systems might fare similarly if not properly assessed.

\subsubsection*{Lack of intellectual freedom} 
Rightly or wrongly, many data scientists currently see ethics requirements as constraints, not just algorithmically but also intellectually. Many see their purpose as discouraging work that does not adhere to certain ``ethical'' standards. Such rhetoric can engender a defensive attitude towards those demanding these standards for fear of one's ideas being shunned and suppressed. Fostering a defensive research culture could have a chilling effect on data science progress, and in some cases even lead to ``black market'' research being done away from ethicists' eyes. Ideally, when practitioners encounter ethical dilemmas, they should feel comfortable reporting and debating them with the larger research community, not suppressing them for fear of ridicule or punishment.

\subsubsection*{Limiting progress for an application/problem} In 2004, a chess program called Fruit \cite{fruit} was open sourced, which had a much lower playing strength than top human grandmasters at the time. However, it led to a revolution where ideas encoded in this initial version were imbibed in other programs and improved upon leading to the current generation of programs such as Stockfish. These programs are nearly unbeatable by humans and are now commonly used by top players for training and in the discovery of new ideas. This was a clear case where imperfect versions can still lead to significant progress. Refinement can happen over time. Analogously in AI, good ideas could be killed early if too many ethical restrictions are imposed upfront.

\subsubsection*{Limiting progress across applications/problems} Likewise, some ideas or tricks used to develop one system or application are often useful for different applications. For example, transformer architectures \cite{transformer} were initially shown to be successful in natural language processing tasks, however, recently they are now preferred even in computer vision tasks \cite{vt}. Hence, the organic transfer of ideas can have impacts far beyond the original application. Yet this intellectual transfer may be severely curtailed if strict yet subjective constraints are placed on what is ethically acceptable.

\subsubsection*{Subjectivity of ethics standards}
Values can be highly subjective at an individual level and even at the level of regulatory institutions. For instance, the requirements of the General Data Protection Regulation (GDPR)  \cite{gdpr} in Europe vary from those outlined by NIST \cite{nist} in the United States. Values are also hard to precisely specify and formalize. This can lead to frustration for persons in academia and industry alike. Publishing in respected venues is already challenging enough. A PhD student trying to meet publication requirements could get frustrated with additional and vague requirements that resemble a moving target. Similar ambiguities could betide engineering teams in industry offering products and services. But subjectivity is not the same as arbitrariness. Robust ethical guidelines and principles drawn from biomedical fields might be adapted for AI research (though this is easier said than done\cite{mittelstadt2019principles}).

\subsubsection*{Neutrality of most technology} A car could be used to transport an ailing person to the hospital  and saving their life, or run over someone. Analogously in AI, a fairness algorithm could be used to output unfair results by misspecifying protected attributes. In line with the atomist's commitment to value-neutrality, technology can be used for good or bad and may not be inherently harmful. Hence, putting restrictions on it might make little sense. Rather, applications should be monitored and constrained.

\section{Targeted Recipes for More Productive Interdisciplinary AI Ethics Discussions}
\label{AccEmp}

\begin{figure*}[h]
\centering
\includegraphics[width=.8\textwidth]{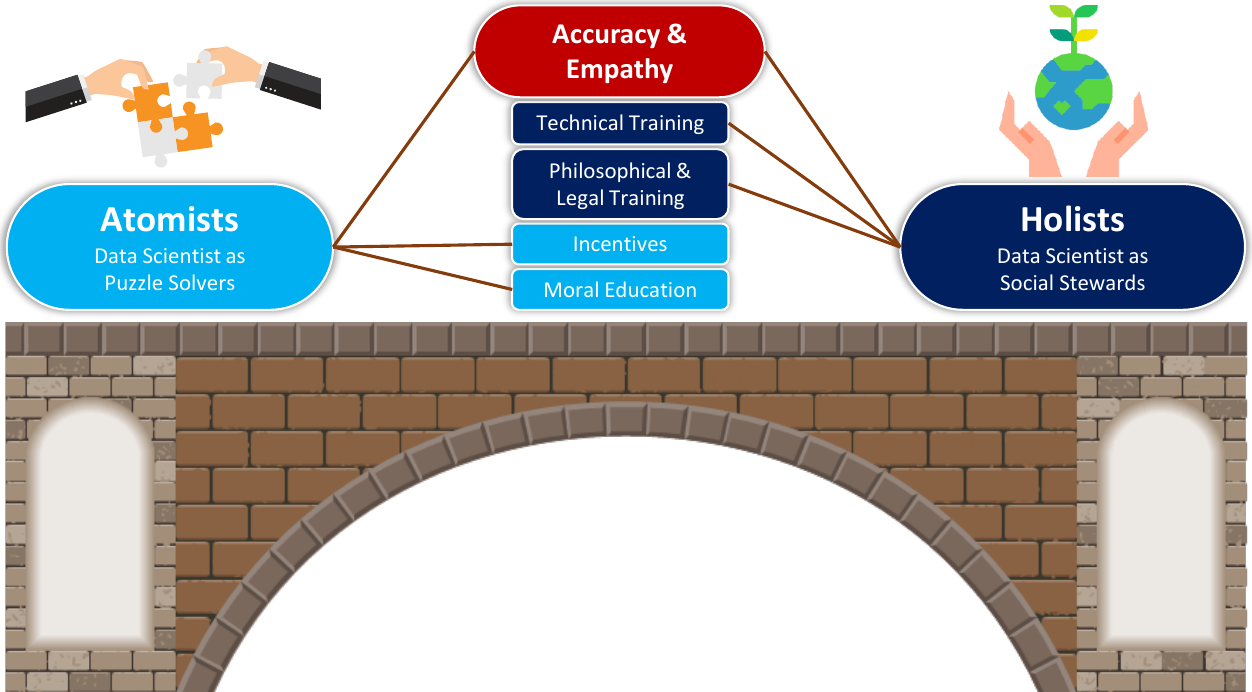}
\caption{The data science community risks splitting into two opposing factions---\emph{atomists} and \emph{holists}---over contentious AI ethics issues. To promote more accurate and empathetic AI ethics discussions, we propose four discipline-targeted recipes to bridge these ideological divisions, reduce community polarization, and ensure AI research benefits society. }
  \label{fig:summary}
\end{figure*}

Given the ideological differences of atomists and holists, we adapt a therapeutic technique to restore trust between them. Guided by twin themes of greater accuracy and empathy (see Figure \ref{fig:summary}), we then present four basic recipes---targeted at different data science community member backgrounds---for fostering more productive dialogue on AI ethics issues to benefit society (see Table \ref{tab:my-tableMat}). 

\subsection{A therapeutic model for interdisciplinary accuracy and empathy}

\begin{table}[b]
\caption{Examples of data science community member backgrounds and positions in the atomist-holist taxonomy. Note that ideological assumptions can be implicit or explicit, largely cutting across disciplinary divides.}  
\label{tab:my-tableMat}
\begin{tabular}{|l|l|l|}
\hline
 &
  \textbf{Atomist} &
  \textbf{Holist} \\ \hline
\textbf{Implicit} &
  \begin{tabular}[c]{@{}l@{}}Engineering, Physics, \\ Computer Science, Statistics\end{tabular} &
  Technology Activism \\ \hline
\textbf{Explicit} &
  \begin{tabular}[c]{@{}l@{}}Economics, \\ Quantitative Social Science\end{tabular} &
  \begin{tabular}[c]{@{}l@{}}Humanities, Law, \\ Qualitative Social Science\end{tabular} \\ \hline
\end{tabular}
\end{table}

Carl Rogers in 1951 proposed a \emph{person-centered approach} to psychotherapy useful for resolving interpersonal conflict \cite{accemp}. Importantly, dialogue participants are more satisfied and less resentful regardless of outcome. A similar dialogical process may be useful in improving the quality of ethics discussions in general. The core idea is the following: \\

\noindent\textit{Accuracy:} Alice first expresses her worry. Bob then repeats it as accurately as possible. This repetition should not be an interpretation of what Bob thinks Alice meant, but as accurate a reproduction as possible. Afterwards, Bob asks Alice if he missed something, and if so, tries again until Alice is satisfied.

\noindent\textit{Empathy:} Once Alice is satisfied with Bob's reformulation, Bob empathizes with her concern. He imagines Alice's perspective and follows her line of reasoning, although he may not agree with it. Then Bob makes his point, which now Alice must accurately repeat. The process continues until both parties are satisfied. A fictional conversation might proceed as follows:

\begin{itemize}\setlength\itemsep{0em}
\item\textit{Alice:} I spent the last few weeks testing Schmoogle's newest large language model for biases. I am now convinced it might be sentient.

\item\textit{Bob:} You said you ``spent the last few weeks testing Schmoogle's newest large language model for biases and now you are convinced it is sentient.'' Is that right?

\item\textit{Alice:} Yes, that is right. We might want to ask for consent before we continue running A/B tests on it.

\item\textit{Bob:} I understand why you would think such a complex language model might be sentient. If I spent as much time as you testing it, I would also be concerned. But the model is just finding statistical correlations in massive amounts of text from all over the internet.

\item\textit{Alice:} You said, ``The model is just finding statistical correlations in massive amounts of text data from all over the internet'' Is that right?

\item \textit{Bob:} $\cdots$
\end{itemize}


\subsection{Moral education for data scientists}
 Moral education is about the socialization and development of virtuous persons: persons with ethically admirable character traits and dispositions who empathize with and care for particular others in their community \cite{noddings2013caring}.  Today, however, most data scientists receive a highly technical education. Skills, not virtues---excellences that contribute to a flourishing human life and community---are the focus. 
 Understandably, data science graduates may implicitly apply their disciplinary standards of objectivity to ethical issues, realize it cannot easily be done, and conclude that anything goes, or that ethical judgments are meaningless. Yet neglecting the moral education of data scientists can leave open a moral vacuum liable to be filled by tribalism, dogmatic identity politics, relativism, or nihilism. Data science educators would do well to emphasize the history and arguments behind the rejection of the fact-value distinction \cite{putnam2004collapse}, as these points seem to have gone unnoticed in AI-feeder disciplines.

Greater debate about the ideal character and virtues of data scientists can also pave the way for professionalization. Example debate-starters might be: under what sort of conditions is corporate whistleblowing appropriate? What is the difference between ethics consulting and fairwashing? To what extent should data scientists be expected to reasonably foresee the harmful use of AI-based technologies by others? While we cannot speak for the community, at the very least the moral education of future data scientists should aim at cultivating an ability to empathize, tolerate, and negotiate in good faith with those whose views differ from their own, without resorting to name-calling and threats of cancellation.

\subsection{Incentives for economists}
To encourage atomists to cooperate with holists to tackle the complex ethical issues posed by AI-based technologies, one suggestion draws on the old idea that commerce and morality are mutually reinforcing \cite{hirschman1982rival}. By providing economic incentives towards more civil engagements on AI ethics issues, we can also respect atomists' desire for autonomy and personal integrity. 

At the institutional level, academic data science departments might incentivize faculty to explore co-teaching data science ethics courses with philosophers and lawyers. Senior researchers and practitioners could also be incentivized to mentor junior data scientists, focusing on character development. Industry data scientists could be rewarded for convening ``ethics roundtable'' discussions or ethics reading groups. Conferences and journals with ethics reviews and impact statements could also offer prizes or community recognition to authors whose statements are particularly cogent or creative \cite{prunkl2021institutionalizing}.  Similarly, positive examples of civil debate---whether on social media or offline---could be rewarded through community praise. The actions of certain ``moral exemplars'' can become teaching material for data science ethics courses.  In general, organizations can do more to frame ethics issues as new and exciting research challenges, rather than as autonomy-restricting handcuffs.

\subsection{Philosophical and legal training for technology activists}
While typically trained in the natural sciences, technology activists seek to draw attention to social justice issues related to AI and data science, such as racial and gender discrimination in STEM fields and industry hiring practices. Technology activists author popular books \cite{o2016weapons, broussard2018artificial} and have  received major media attention in news articles \cite{wylieCambridgeAn, mitchellFired} and films (i.e., \emph{The Social Dilemma}). Although we sympathize with activists' intentions, we also worry that threatening the cancellation\cite{animacancellation} of those who hold opposing values can stoke fears of tribalism and dogmatic group-think, sparking further disciplinary polarization. Indeed, given activists' concern with the unjust use of power, threats of mob-based cancellation may appear hypocritical.

\tg{Echoing similar proposals from science and technology studies and digital humanities scholars \cite{berry2012understanding, riley2008engineering},} we encourage philosophical training to hone technology activists' persuasive and reflective abilities and create new allies and collaborators in the humanities, social sciences, and law. With broad support from the often-fragmented ``two cultures'' of the natural and human sciences, technology activists can arguably better achieve their ethical goals. Philosophy provides an array of concepts to express ethical concerns (e.g., rights, duties, utility, virtue, power, care, etc.) and can aid in identifying \emph{ad hominem} attacks and logical fallacies. Familiarity with various areas of law (e.g., corporate law, public law, torts, etc.) can also help technology activists frame AI harms in terms of pre-existing legal frameworks and concepts, such as discrimination or human rights law \cite{aizenberg2020designing}.

\subsection{Technical training for humanities scholars}
For those in the humanities (e.g., philosophy, law, media and communication studies, etc.) and ``soft'' social sciences (e.g., anthropology, sociology, education, etc.,) who might participate in ethics reviews, we think it fair to ask whether they have the ``appropriate'' technical expertise to accurately evaluate data science products, systems, and/or papers. For instance, there is a growing literature that draws on postmodern philosophical sources to explain how AI can entrench existing racial and gender inequities \cite{benjamin2019race} and reproduce the dynamics of colonial exploitation and power differentials \cite{crawford2021atlas, couldry2019data}. While we welcome critical voices into the debate on the ethics of AI, we must clearly distinguish \emph{features} of the technology in itself from \emph{uses} of the technology by organizations embedded in larger social, political, and economic systems. 

 One suggestion is to establish an accreditation process for ethics reviewers, such as a minimum-duration internship in industry or an AI-related academic department (e.g., computer science or statistics). The knowledge and skills obtained could improve both the accuracy and empathy of cross-disciplinary dialogue. Real-world experience would help ethics reviewers to develop their \emph{practical} moral knowledge and increase their ability to empathize with data scientists. Accreditation could be done in collaboration with a number of interdisciplinary research groups and industry research labs.

\section{Conclusion}
\label{conc}
The data science community appears split---sometimes violently so---along an ideological divide. We explicated key beliefs and assumptions of atomist and holist ideologies and described their relevance to the fact-value debate. Atomists see themselves as autonomous puzzle solvers and hold that facts are and should be kept separate from values, while holists see themselves as social stewards and believe facts and values mutually reinforce one another. In the interests of the data science community and society, we advocate for a balance between the two camps' views and offer several recipes for more empathetic and productive interdisciplinary dialogue on AI ethics. This work can serve as a microcosm of such dialogue, as we ourselves occupy various points along the atomist-holist spectrum.

But our diagnosis and proposal has limitations. First, some recipes may not translate well to digital and social media contexts, where the most-heated AI ethics arguments tend to occur.  Second, the atomist-holist taxonomy obscures heterogeneity among data scientists who may hold less traditional views. Lastly, because many advances in AI happen inside technology corporations, corporations arguably should assume the burdens of legal and moral responsibility for the harms and risks imposed by their innovations. More systemic solutions---at the level of corporate law and governance \cite{stout2012shareholder}---are likely needed to ensure that AI-based technology is developed and deployed with the broader interests of society in mind, not simply shareholders.

 We hope our diagnosis and vision draws attention to the importance of greater empathy, openness, and humility for all members of the data science community, a community whose boundaries  continue to evolve and expand. We look forward to a time when community ethics discussions are dominated by the best reasons, rather than the loudest voices.
 
 \section*{Acknowledgements}
We thank the Editor and three anonymous reviewers for their helpful comments and criticisms, along with Finale Doshi-Velez for feedback on an earlier draft. We also thank David Martens and the Applied Data Mining Research Group at the University of Antwerp for many stimulating conversations. T.G. and G.S. were partially funded by the Taiwan National Science and Technology Council [Grant 111-2410-H-007-030-MY3].

\section*{Author contributions}
  T.G. and A.D. conceived of and wrote the manuscript; G.S. provided supervision and feedback.

\bibliography{AccuracyEmpathyBiblio}

\end{document}